\documentclass[draftclsnofoot, 12pt, onecolumn]{IEEEtran}

\usepackage{graphicx}
\usepackage{color}
\usepackage{placeins}
\usepackage{float}
\usepackage{tabularx,colortbl}
\usepackage{amsmath}
\usepackage{multirow}
\usepackage{hyperref}

\begin{document}
%
\title{Rate Gain Region and Design Tradeoffs for Full-Duplex Wireless Communications}

\author{Elsayed Ahmed, Ahmed Eltawil and Ashutosh Sabharwal\footnote{Elsayed Ahmed and Ahmed Eltawil are with the Department of Electrical Engineering and Computer Science at the University of California, Irvine, CA, USA (e-mail: \{ahmede, aeltawil\}@uci.edu). Ashutosh Sabharwal is with the Department of Electrical and Computer Engineering, Rice University, Houston, TX, USA (e-mail: ashu@rice.edu).}
}

\maketitle

\begin{abstract}
In this paper, we analytically study the regime in which practical full-duplex systems can achieve larger rates than an equivalent half-duplex systems. The key challenge in practical full-duplex systems is uncancelled self-interference signal, which is caused by a combination of hardware and implementation imperfections. Thus, we first present a signal model which captures the effect of significant impairments such as oscillator phase noise, low-noise amplifier noise figure, mixer noise, and analog-to-digital converter quantization noise. Using the detailed signal model, we study the rate gain region, which is defined as the region of received signal-of-interest signal strength where full-duplex systems outperform half-duplex systems in terms of achievable rate. The rate gain region is derived as a piecewise linear approximation in log-domain, and numerical results show that the approximation closely matches the exact region.  Our analysis shows that when phase noise dominates mixer and quantization noise, full-duplex systems can use either active analog cancellation or baseband digital cancellation to achieve near-identical rate gain regions. Finally, as a design example, we numerically investigate the full-duplex system performance and rate gain region in typical indoor environments for practical wireless applications.
\end{abstract}


%
\IEEEpeerreviewmaketitle

\section{Introduction}


One major shortcoming of current deployed systems is the limitation to operate as half-duplex systems employing either a time-division or frequency-division to achieve bidirectional communication. 
The main challenge in full-duplex transmission is due to the large power differential between the self-interference signal caused by the node's own transmission and the signal-of-interest which the receiver intends to decode.  
A number of recent publications~\cite{Ref3}-\cite{Ref9r} have demonstrated experimentally that practical full-duplex systems can improve upon half-duplex systems under certain conditions and using the appropriate self-interference cancellation schemes. However, none of the current full-duplex wireless systems gain over half-duplex transmission in all operation regimes, largely due to imperfect self-interference cancellation. This paper is inspired by the above observation and aims to analytically understand the bottlenecks in full-duplex operation.  

To make progress in understanding the limits of practical full-duplex systems, we use a signal model for narrowband full- and half-duplex systems by modeling four significant transceiver noise sources: (i) transmitter and receiver phase noise, (ii) Low Noise Amplifier (LNA) noise figure, (iii) mixer noise figure, and (iv) Analog-to-Digital Converter (ADC) quantization noise. We use the detailed signal model, with the above mentioned four noise sources, along with different self-interference cancellation mechanisms to analytically investigate the operation regions in which full-duplex systems outperform half-duplex systems in terms of achievable rate, under the same operating conditions.

Throughout the literature, several self-interference cancellation mechanisms have been proposed. Self-interference cancellation mechanisms can be divided into two main categories: (i) passive suppression, and (ii) active cancellation. In passive suppression, the self-interference signal is suppressed in the propagation domain before it goes through the receiver circuitry. Passive suppression could be achieved using antenna separation and/or shielding~\cite{Ref3,Ref5}, directional antennas~\cite{Ref8,Ref9}, or careful antenna placement~\cite{Ref6}. In active cancellation schemes~\cite{Ref4,Ref5}, the self-interference signal is canceled by leveraging the fact that the transceiver knows the signal it is transmitting. Active cancellation schemes can be divided into digital and analog cancellation based on the signal domain (digital-domain or analog-domain) where the self-interference is actively canceled.

In addition to spectral efficiency improvement, several other advantages of full-duplex communication have been investigated in the literature. More specifically, in a cognitive radio context~\cite{Ref9,Ref9rr}, full-duplex communication could be used to avoid interfering to the primary users, by enabling simultaneous transmit and spectrum sensing at the secondary users. Full-duplex communication could also be used to eliminate the hidden terminal problem in wireless networks~\cite{Ref7}. Full-duplex relaying is another application for full-duplex communication node~\cite{Ref10}-\cite{Ref10rr}.

In this paper, we consider the problem of bidirectional full-duplex transmission under both passive suppression and active cancellation schemes for the following results.

First, we derive the full-duplex rate gain region under different operating conditions. We define the rate gain region as the region of received signal-of-interest strength at which full-duplex systems outperform half-duplex systems in terms of achievable rate. Our key contribution is a closed-form piecewise linear approximation, in the log-domain, for the rate gain region under analog and digital self-interference cancellation schemes. In the log-domain means that the relation is between the logarithm of the variables, e.g. a linear relation between $y$ and $x$ in the log-domain means that $\log(y)=\log(x)+\log(constant)$. The piecewise linear approximation allows us to develop valuable insights in the behavior of full-duplex systems under different operating conditions. Second, we identify the dominant noise components that limit system performance for both analog and digital self-interference cancellation schemes.

Third, we investigate the possible design tradeoffs involved in full-duplex system design. The results show that, the rate gain region is inversely proportional to the transmit power, i.e. reducing the transmit power makes full-duplex systems more likely to outperform half-duplex systems, which is consistent with the results in~\cite{Ref10r}. Despite dealing with different system architecture (relay systems), the work in~\cite{Ref10r} show that, as the signal-to-noise ratio decreases (i.e. lower transmit power), full-duplex relay is more likely to outperform half-duplex relay in terms of outage probability.

Finally, the paper concludes with a design example, where we quantify the design requirements to enable full-duplex transmission with rate gains as compared to half-duplex transmission. The results show that, for typical indoor environments, a $-$60dBc in-band phase noise combined with 40dB passive suppression could achieve a full-duplex rate of $\sim$1.2x to $\sim$1.4x times that of half-duplex systems. Although the results show a considerable improvement in the system achievable rate, it also highlights the design challenges represented by the strong passive self-interference suppression and the low-noise restrictions.

We note that the choice of impairments to model, and the accuracy with which they are modeled has a significant impact on the overall model reliability. For example, the models presented in~\cite{Ref3,Ref4,Ref9} account for the transmitter phase noise and the receiver thermal noise. However, under certain conditions, other noise factors dominate performance. For instance, for large self-interference signal power, the ADC quantization noise becomes a performance limiting factor, especially when the number of ADC bits are small. Furthermore, the LNA is forced to operate in a lower gain region, which implies that the overall noise figure is no longer dominated by the LNA noise figure, but rather it becomes important to distinguish noise sources and dependencies. Another work in~\cite{Ref11} develops a signal model for a full-duplex multi-input multi-output system combining all radio impairments together in one system parameter. The proposed approach in~\cite{Ref10rr,Ref11} simplifies system analysis but results in a loss of modeling fidelity since it becomes difficult to associate specific components with performance loss. 

The remainder of this paper is organized as follows. In Section II, the signal model is presented. The rate gain region is derived in Section III. System behavior and design tradeoffs are discussed in Section IV. Section V presents the numerical results. And finally, section VI presents the conclusion.

\section{Signal Model}
\begin{figure}[!ht]
\begin{center}
\noindent
  \includegraphics[width=3.5in, trim= 0in 0in 0in 0in]{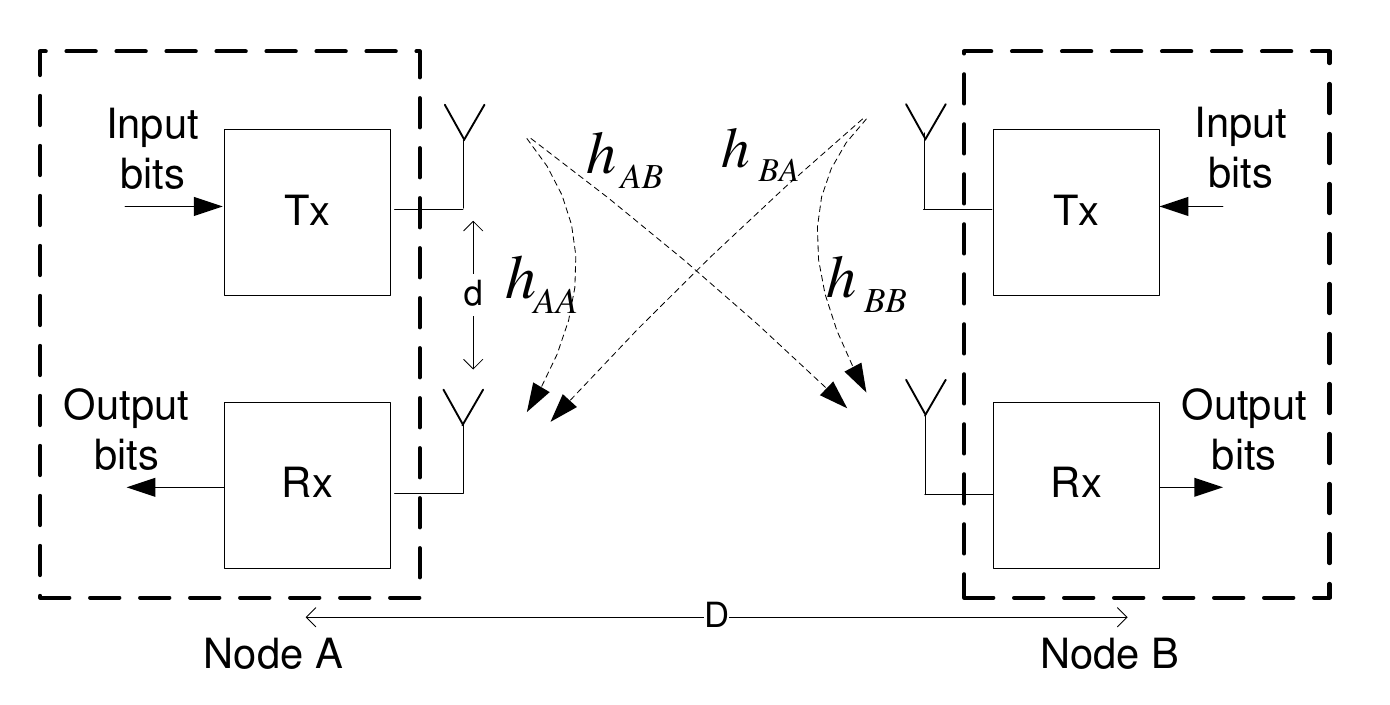}
  \caption{Single-input single-output full-duplex system.\label{Fig1Label}}
\end{center}
\end{figure}

\begin{figure}[!ht]
\begin{center}
\noindent
  \includegraphics[width=5in, trim= 0in 0in 0in 0in]{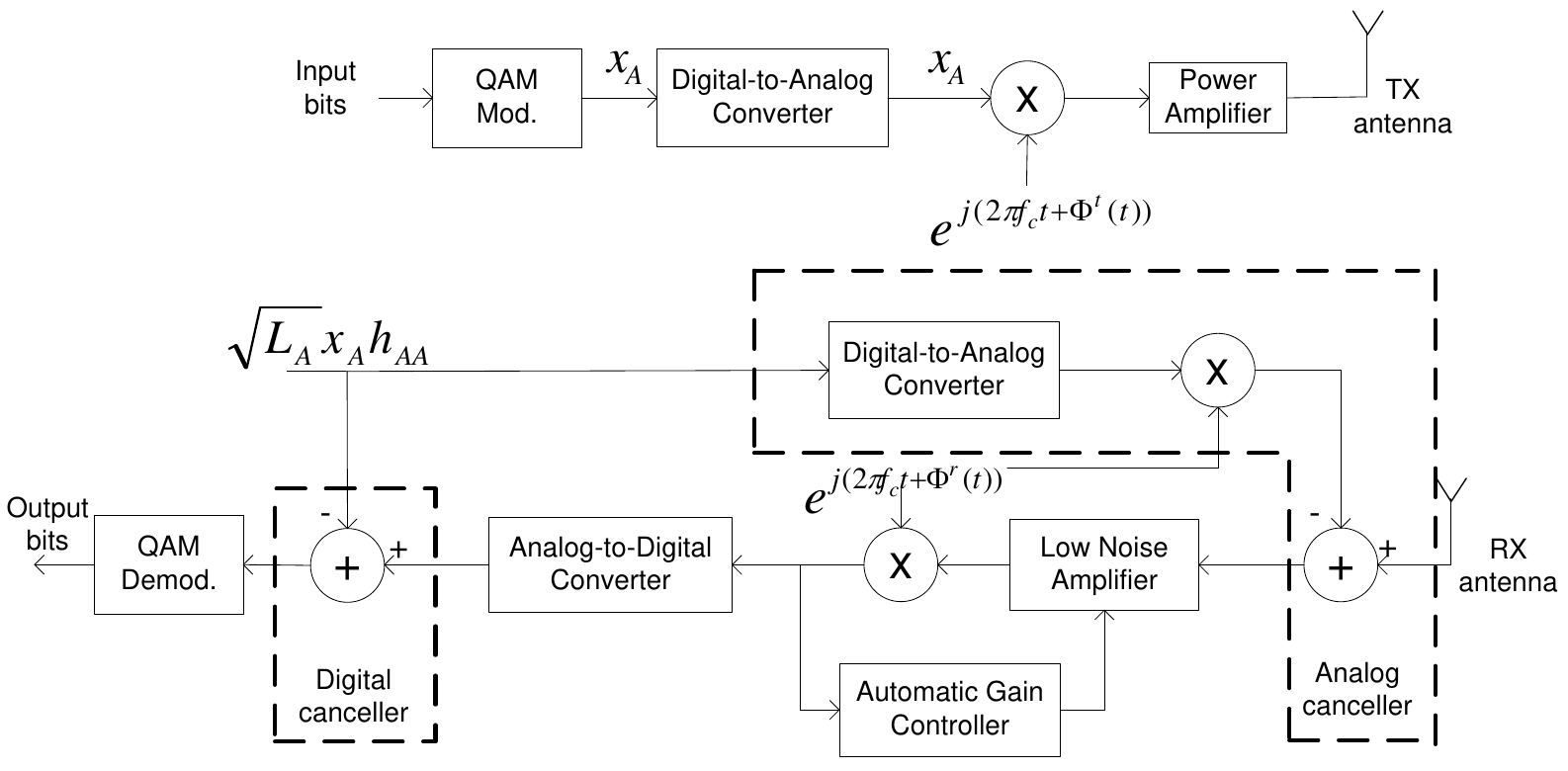}
  \caption{Detailed block diagram of a full-duplex transceiver with analog or digital self-interference cancellation.\label{Fig2Label}}
\end{center}
\end{figure}

In this section, we describe the signal models for both half and full-duplex narrowband systems. As illustrated in Figures~\ref{Fig1Label} and~\ref{Fig2Label}, node A and B are separated by distance $D$ meter, and communicating in a full-duplex manner. For conceptual clarity, we have depicted the transmit antenna separate from the receive antenna in Figures~\ref{Fig1Label} and~\ref{Fig2Label}. However, our model and subsequent analysis applies to circulator-based systems which use only one antenna for both transmission and reception. At the transmitter side, the signal is modulated and then up-converted to the carrier frequency $f_c$, the oscillator at the transmitter side is assumed to have a random phase error represented by $\phi^t(t)$. The signal is then amplified by the transmitter power amplifier. 

At the receiver side, the incoming signal level is appropriately adjusted using the LNA which is controlled by the automatic gain control block. The signal is then down-converted from the carrier frequency, the down-conversion mixer is assumed to have a random phase error represented by $\phi^r(t)$. The down-converted signal is then quantized using an $m$-bits ADC.

For simplicity and without loss of generality, we consider the signal at node A, where, due to hardware symmetry, the same analysis applies to node B. According to the described model, the received signal \emph{without} self-interference cancellation can be written as
\begin{equation}\label{eq:1}
y[n]= \left( \sqrt{L_A} x_A[n] h_{AA}[n] e^{i\phi_A^t[n]} + \sqrt{L_B} x_B[n] h_{BA}[n] e^{i\phi_B^t[n]} \right)e^{i\phi_A^r[n]} + z[n]+ q[n]\text{,}
\end{equation}
where $x_A$, $x_B$ are the transmitted signal from node A and B, $h_{AA}$, $h_{BA}$ are the self-interference and signal-of-interest channels respectively, $L_A$, $L_B$ are the propagation losses due to the antenna isolation at the same node and the distance between the two communicating nodes respectively, $\phi_i^t, \phi_i^r$, $i \in[A,B]$ are the carrier phase error at the transmitter and receiver side of node $i$, $z$ is the receiver noise, and $q \sim \mathcal{U}(0,\frac{1}{2^{m-1}})$ is the uniformly distributed ADC quantization noise, where $m$ is the number of ADC bits.

The  receiver noise, $z[n]$, represents the additive noise inherent in the receiver circuits, and specified by the circuit noise figure. The overall receiver noise power can be calculated as~\cite{Ref13}
\begin{equation}\label{eq:7}
P_z = P_{th} N_f = P_{th} \left(N_l+\frac{N_m-1}{\alpha^2}\right)\text{,}
\end{equation}
where $N_f$ is the overall receiver noise figure, $N_l$ is the LNA noise figure, $N_m$ is the mixer noise figure, $P_{th}$ is the thermal noise power in a 50ohm source resistance, and $\alpha^2$ is the LNA power gain. The ADC quantization noise is a uniformly distributed noise introduced by the ADC due to the signal quantization. For an $m$ bits ADC, the total ADC quantization noise power is calculated in terms of the LNA power gain as~\cite{Ref14}
\begin{equation}\label{eq:10}
P_q = \frac{1}{\alpha^2} \frac{1}{12 \cdotp 2^{2m-2}}=\frac{\sigma_q^2}{\alpha^2}\text{,}
\end{equation}
where $\sigma_q^2 = \frac{1}{12 \cdotp 2^{2m-2}}$ is the quantization noise variance.

In a full-duplex system, digital or analog self-interference cancellation scheme is used to mitigate the self-interference signal. Both digital and analog cancellation require the knowledge of the transmitted signal and the self-interference channel. In our analysis, the channel is assumed to be frequency-flat fading channel, and the channel state information for all transmitter-receiver links are assumed to be perfectly known at the receiver side. 

The main difference between digital and analog canceler is the signal domain, digital or analog domain, where the self-interference signal is cancelled. In digital cancellation scheme, the interference signal is eliminated in the digital domain after the received signal goes through the radio section, which forces the LNA to operate at low gain modes. However, in analog cancellation scheme, the base-band self-interference signal is up-converted to the carrier frequency and then subtracted from the received signal in the analog domain. The elimination of the self-interference signal in the analog domain, before the received signal goes through the LNA, allows the LNA to operate at higher gain than in digital cancellation scheme. 

\subsection{Digital cancellation scheme}
In digital cancellation scheme, with the knowledge of the interference channel, the self-interference cancellation is done by subtracting the signal $\sqrt{L_A} x_A[n] h_{AA}[n]$ from the received signal in the digital domain. After digital self-interference cancellation the remaining signal can be written as
\begin{equation}\label{eq:2}
y[n]=\sqrt{L_A} x_A[n] h_{AA}[n] \left(e^{i(\phi_A^t[n]+\phi_A^r[n])}-1 \right)+\sqrt{L_B} x_B[n] h_{BA}[n] e^{i(\phi_B^t[n]+\phi_A^r[n])}+ z[n]+ q[n]\text{.}
\end{equation}
Using the approximation of $e^{i\phi}\cong 1+i\phi$ for $\phi\ll 1$, Equation~\eqref{eq:2} can be written as
\begin{equation}\label{eq:3}
y[n]=\sqrt{L_A} x_A[n] h_{AA}[n] \left(i\phi_A^t[n]+ i\phi_A^r[n]\right)+\sqrt{L_B} x_B[n] h_{BA}[n] \left(1+i\phi_B^t[n]+ i\phi_A^r[n]\right) + z[n]+ q[n]\text{.}
\end{equation}

The resulting full-duplex Signal-to-Interference plus Noise Ratio (SINR) for digital cancellation scheme can be written as
\begin{equation}\label{eq:4}
\mathsf{SINR}^{\mathsf{FD}}_{\mathsf{DC}} = \frac{P_x L_B |h_{BA}|^2}{P_{\phi,A}+P_{\phi,B}+P_{z,\mathsf{DC}}+P_{q,\mathsf{DC}}}\text{,}
\end{equation}
where acronyms $\mathsf{DC}$ and $\mathsf{FD}$ refers to digital cancellation and full-duplex, respectively. Further, $P_x=E\{|x_A|^2\}=E\{|x_B|^2\}$ is the transmitted signal power, $E\{\}$ denotes expectation process, and $P_{\phi,A}$, $P_{\phi,B}$ are the self-interference and signal-of-interest phase noise power calculated as
\begin{equation}\label{eq:5}
P_{\phi,A}=P_x L_A |h_{AA}|^2 \left(\mu_A^t+\mu_A^r\right)=P_x L_A |h_{AA}|^2 \mu\text{,}
\end{equation}
\begin{equation}\label{eq:6}
P_{\phi,B}=P_x L_B |h_{BA}|^2 \left(\mu_B^t+\mu_A^r\right)=P_x L_B |h_{BA}|^2 \mu \text{,}
\end{equation}
where $\mu_i^t$, $\mu_i^r$, $i \in[A,B]$ are the total transmitter and receiver normalized phase noise power. In this paper we assume that node A and B are hardware symmetrical. Therefore, the \emph{statistics} of the transmitter phase noise are identical, thus $\mu_A^t = \mu_B^t$. The total phase noise power ($\mu_i^j$, $i\in[A,B]$, $j\in[t,r]$) is calculated by integrating the power spectral density (PSD) of the corresponding phase noise process ($\phi_i^j$) over the system's bandwidth. Generally, the PSD is a design dependent parameter that depends on the architecture and design parameters of the used phase-locked loop (PLL). Approximated expressions of the phase noise's PSD for different PLL designs could be obtained~\cite{Ref12}. However, precise phase noise PSD is usually obtained through measurements of a fabricated tuner involving a PLL.

The receiver and quantization noise power for digital cancellation scheme ($P_{z,\mathsf{DC}}$, $P_{q,\mathsf{DC}}$) are calculated in terms of the LNA power gain as in~\eqref{eq:7} and~\eqref{eq:10}. The LNA power gain is calculated by the variable gain amplifier circuit in terms of the received signal power at the LNA input as
\begin{equation}\label{eq:8}
\alpha_{\mathsf{DC}}^2 = \frac{1}{P_x L_A |h_{AA}|^2 + P_x L_B |h_{BA}|^2}\text{.}
\end{equation}

Define the instantaneous received signal strength (RSSI) for the self-interference and signal-of-interest respectively as $\mathsf{RSSI}_A = P_x L_A |h_{AA}|^2$, $\mathsf{RSSI}_B = P_x L_B |h_{BA}|^2$, then using~\eqref{eq:7},~\eqref{eq:10},~\eqref{eq:5},~\eqref{eq:6} and~\eqref{eq:8} in~\eqref{eq:4}, we get
\begin{equation}\label{eq:11}
\mathsf{SINR}_{\mathsf{DC}}^{\mathsf{FD}} = \frac{\mathsf{RSSI}_B}{\eta \mathsf{RSSI}_A + \eta \mathsf{RSSI}_B + \zeta}\text{,}
\end{equation}
where
\begin{equation}\label{eq:13}
\eta = \mu + \sigma_q^2 + P_{th} N_m - P_{th}\text{,}
\end{equation}
\begin{equation}\label{eq:14}
\zeta = P_{th} N_l\text{.}
\end{equation}
The parameter $\zeta$ can be described as the \emph{system noise power floor} that does not depend on the incoming signal power. On the other hand, the parameter $\eta$ represents the \emph{signal power dependent noise component}.
\subsection{Analog cancellation scheme}
In some of the active analog cancellation schemes, the self-interference cancellation is done by subtracting the up-converted self-interference signal from the received signal in the analog domain, before the received signal goes through the LNA~\cite{Ref5}. The residual signal after analog self-interference cancellation can be written as
\begin{equation}\label{eq:2r}
y[n]=\sqrt{L_A} x_A[n] h_{AA}[n] \left(e^{i\phi_A^t[n]} - e^{i\phi_A^r[n]}\right) e^{i\phi_A^r[n]} +\sqrt{L_B} x_B[n] h_{BA}[n] \text{\ }e^{i\phi_B^t[n]}e^{i\phi_A^r[n]}+ z[n]+ q[n]\text{.}
\end{equation}
Using the approximation of $e^{i\phi}\cong 1+i\phi$ for $\phi\ll 1$ and collecting terms, Equation~\eqref{eq:2r} can be written as
\begin{equation}\label{eq:3r}
y[n]=\sqrt{L_A} x_A[n] h_{AA}[n]\left(i\phi_A^t[n]-i\phi_A^r[n]\right) +\sqrt{L_B} x_B[n] h_{BA}[n] \left(1+i\phi_B^t[n]+i\phi_A^r[n]\right) + z[n]+ q[n]\text{.}
\end{equation}
The resulting SINR for analog cancellation scheme can be written as
\begin{equation}\label{eq:4r}
\mathsf{SINR}^{\mathsf{FD}}_{\mathsf{AC}} = \frac{P_x L_B |h_{BA}|^2}{P_{\phi,A}+P_{\phi,B}+P_{z,\mathsf{AC}}+P_{q,\mathsf{AC}}}\text{,}
\end{equation}
where the acronym $\mathsf{AC}$ refers to analog cancellation.

Comparing~\eqref{eq:4} and~\eqref{eq:4r}, it has to be noticed that the only difference between the SINR in digital and analog cancellation schemes is the value of the receiver and quantization noise power. Generally, the receiver and quantization noise power are inversely proportional to the LNA power gain (review~\eqref{eq:7} and~\eqref{eq:10}). In analog cancellation scheme the self-interference signal is eliminated before the signal goes through the LNA, which forces the LNA to operate at high gain mode, and thus reducing the receiver and quantization noise power. The LNA power gain for analog cancelation scheme is calculated as
\begin{equation}\label{eq:9}
\alpha_{\mathsf{AC}}^2 = \frac{1}{P_x L_B |h_{BA}|^2}\text{.}
\end{equation}
Substituting from~\eqref{eq:7},~\eqref{eq:10},~\eqref{eq:5},~\eqref{eq:6}, and~\eqref{eq:9} in~\eqref{eq:4r} we get
\begin{equation}\label{eq:12}
\mathsf{SINR}_{\mathsf{AC}}^{\mathsf{FD}} = \frac{\mathsf{RSSI}_B}{\mu \mathsf{RSSI}_A + \eta \mathsf{RSSI}_B + \zeta}\text{.}
\end{equation}

Equation~\eqref{eq:11},~\eqref{eq:12} describes the full-duplex system SINR for both digital and analog cancellation schemes. In our analysis, we compare full-duplex system performance against half-duplex system performance. Half-duplex transmission can be considered a special case of the full-duplex transmission, where the received self-interference signal power is equal to zero and the temporal resources are divided between the two nodes. To maintain a fair comparison, the transmitted power is doubled in the case of half-duplex transmission since only one node is transmitting at a time. Accordingly the half-duplex system Signal-to-Noise Ratio (SNR) can be written as
\begin{equation}\label{eq:15}
\mathsf{SNR}^{\mathsf{HD}} = \frac{2\mathsf{RSSI}_B}{2 \eta \mathsf{RSSI}_B + \zeta}\text{.}
\end{equation}

\section{Rate gain region for digital and analog cancellation schemes}
In this section, we derive the full-duplex rate gain region for both digital and analog cancellation schemes. Rate gain region is defined as the region of received signal-of-interest strength at which full-duplex system achieves rate gain over half-duplex system. Deriving the rate gain region allows for straightforward exploration of the conditions at which full-duplex systems outperform half-duplex counterparts. The rate gain region can be obtained by solving the following inequality
\begin{equation}\label{eq:16}
\mathsf{R}^{\mathsf{FD}} > \mathsf{R}^{\mathsf{HD}}\text{,}
\end{equation}
where $\mathsf{R}^{\mathsf{FD}}$ ,$\mathsf{R}^{\mathsf{HD}}$ are the full-duplex and half-duplex system achievable rates respectively. 

Generally, deriving the rate gain region of a full-duplex system depends on how it is defined. For example, the rate gain region could be defined as the region in which the full-duplex \emph{sum} rate is greater than the half-duplex \emph{sum} rate (i.e. $\mathsf{R}^{\mathsf{FD}}_{A \rightarrow B}+\mathsf{R}^{\mathsf{FD}}_{B \rightarrow A} > \mathsf{R}^{\mathsf{HD}}_{A \rightarrow B}+\mathsf{R}^{\mathsf{HD}}_{B \rightarrow A}$). Although this is the general definition, there might be a scenario where the full-duplex sum rate is better than the half-duplex sum rate, while one of the two communication links has smaller full-duplex rate than its half-duplex rate (i.e. $\mathsf{R}^{\mathsf{FD}}_{A \rightarrow B}+\mathsf{R}^{\mathsf{FD}}_{B \rightarrow A} > \mathsf{R}^{\mathsf{HD}}_{A \rightarrow B}+\mathsf{R}^{\mathsf{HD}}_{B \rightarrow A}$, while $\mathsf{R}^{\mathsf{FD}}_{A \rightarrow B} < \mathsf{R}^{\mathsf{HD}}_{A \rightarrow B}$). In this case, one node has to sacrifice part of its rate, which might not be practical especially in symmetric communication scenarios. Another conservative definition for the rate gain region, is the region in which the full-duplex rate for \emph{each} communication link is greater than its half-duplex rate (i.e. $\mathsf{R}^{\mathsf{FD}}_{A \rightarrow B} > \mathsf{R}^{\mathsf{HD}}_{A \rightarrow B}$ and $\mathsf{R}^{\mathsf{FD}}_{B \rightarrow A} > \mathsf{R}^{\mathsf{HD}}_{B \rightarrow A}$). In this case the full-duplex sum rate is guaranteed to be greater than the half-duplex sum rate.

In this paper we analyze the rate gain region based on the second definition, i.e. the rate gain region is defined as the region in which the full-duplex rate for each communication link is greater than its half-duplex rate. Defining the rate gain region that way makes the analysis applicable for other full-duplex systems' architectures such as in~\cite{Ref8,Ref10,Ref10r}, where only one node (base-station or relay node) is operating in full-duplex mode and communicating with two half-duplex nodes. Accordingly, we derive the rate gain region for one of the two communicating nodes, and the same results apply to the other node, but with different parameters' values. The per-direction achievable rates at node A is calculated as
\begin{equation}\label{eq:17}
\mathsf{R}^{\mathsf{FD}} = \log_2\left(1+\mathsf{SINR}^{\mathsf{FD}}\right)\text{,}
\end{equation}
\begin{equation}\label{eq:18}
\mathsf{R}^{\mathsf{HD}} = \frac{1}{2} \log_2\left(1+\mathsf{SNR}^{\mathsf{HD}}\right)\text{.}
\end{equation}
The factor of $\frac{1}{2}$ is due to the fact that node A and B are sharing the available temporal resources. Substitute from~\eqref{eq:17} and~\eqref{eq:18} in~\eqref{eq:16} we get
\begin{equation}\label{eq:19}
\log_2\left(1+\mathsf{SINR}^{\mathsf{FD}}\right) > \frac{1}{2} \log_2\left(1+\mathsf{SNR}^{\mathsf{HD}}\right)\text{.}
\end{equation}
Equation (19) could be reduced to
\begin{equation}\label{eq:20}
\left(1+\mathsf{SINR}^{\mathsf{FD}}\right)^2 > \left(1+\mathsf{SNR}^{\mathsf{HD}}\right)\text{,}
\end{equation}
then
\begin{equation}\label{eq:21}
\left(\mathsf{SINR}^{\mathsf{FD}}\right)^2 + 2 \mathsf{SINR}^{\mathsf{FD}} > \mathsf{SNR}^{\mathsf{HD}}\text{.}
\end{equation}
In the following analysis, Equation~\eqref{eq:21} along with the signal model presented in Section II are used to derive the rate gain region for both digital and analog cancellation schemes.
\subsection[Part:A]{Rate gain region for digital cancellation scheme \label{sec:digitalrate}}
Substituting from~\eqref{eq:11}, and~\eqref{eq:15} in~\eqref{eq:21} we get
\begin{equation}\label{eq:22}
\left(\frac{\mathsf{RSSI}_B}{\eta \mathsf{RSSI}_A + \eta \mathsf{RSSI}_B + \zeta}\right)^2 + \frac{2 \mathsf{RSSI}_B}{\eta \mathsf{RSSI}_A + \eta \mathsf{RSSI}_B + \zeta} > \frac{2 \mathsf{RSSI}_B}{2 \eta \mathsf{RSSI}_B + \zeta}\text{.}
\end{equation}
Collecting terms and putting~\eqref{eq:22} in the form of a $2^{nd}$ order inequality we get
\begin{equation}\label{eq:23}
a (\mathsf{RSSI}_B)^2 + b \mathsf{RSSI}_B + c > 0\text{,}
\end{equation}
where
\begin{equation}\label{eq:24}
a=\eta(\eta+1)\text{, }b=\zeta(\eta+\frac{1}{2})\text{, } c = -\eta \mathsf{RSSI}_A (\eta \mathsf{RSSI}_A + \zeta)\text{.}
\end{equation}
Knowing that RSSI is always a positive quantity and noting that $c$ is always negative, the rate gain region for digital cancellation scheme can be written as
\begin{equation}\label{eq:27}
\mathsf{RSSI}_B > \mathsf{RSSI}_{B,min}\text{,}
\end{equation}
where
\begin{equation}\label{eq:28}
\mathsf{RSSI}_{B,min} = \frac{-b + \sqrt{b^2-4ac}}{2a}\text{.}
\end{equation}

\begin{figure}[!ht]
\begin{center}
\noindent
  \includegraphics[width=3.5in,trim= 0in 0in 0in 0in]{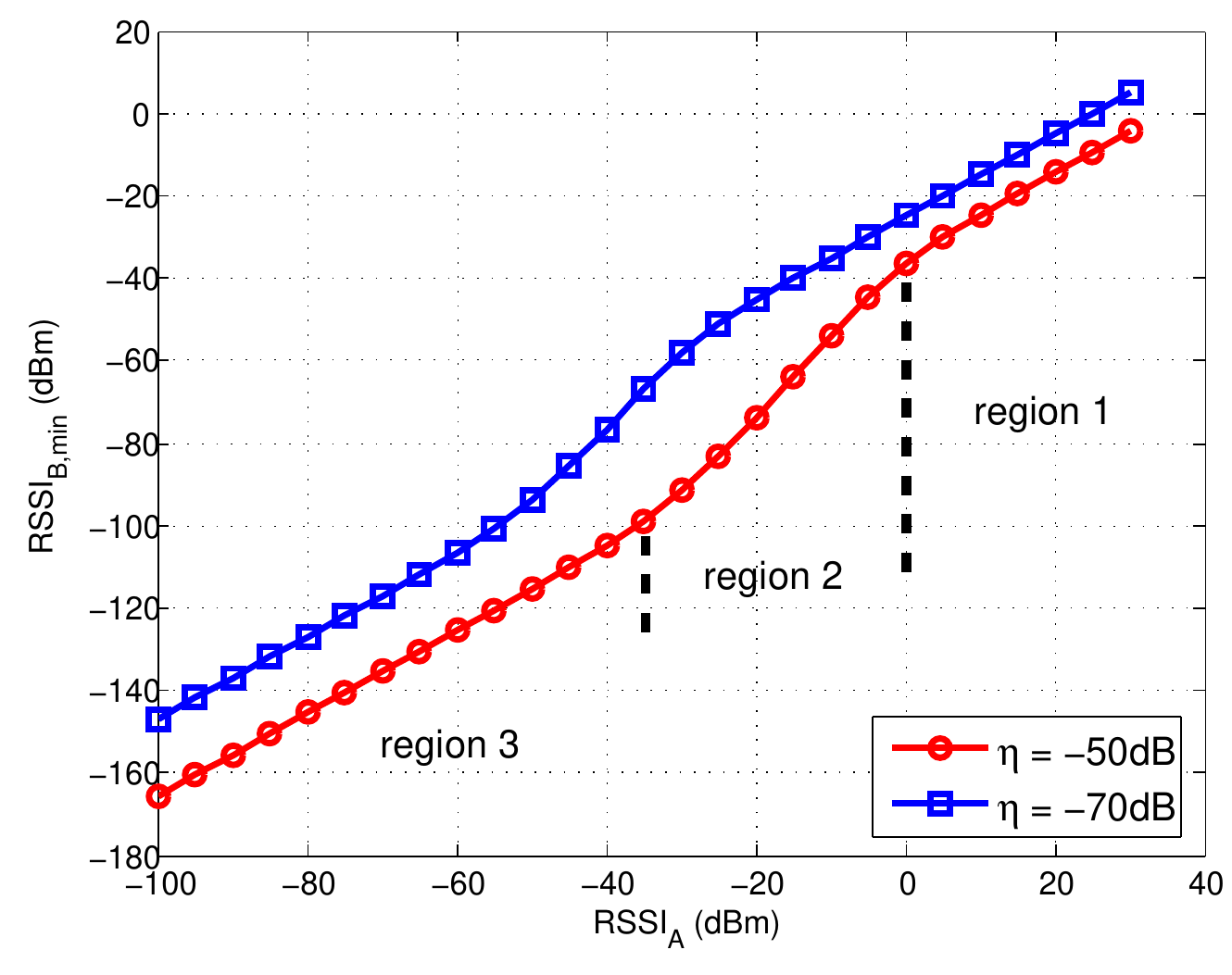}
  \caption{Rate gain region for digital cancellation scheme.\label{Fig3Label}}
\end{center}
\end{figure}

Equation~\eqref{eq:28} describes the rate gain region in terms of all system parameters and radio impairments. However, due to the complexity of the formula in~\eqref{eq:28}, it is difficult to gain insights into system behavior under different operation conditions without resorting to numerical simulations. In the following analysis we try to simplify the relation in~\eqref{eq:28} by deriving a piecewise linear, in the log-domain, approximation for the rate gain region. The results in Figure~\ref{Fig3Label} show that based on the received self-interference signal strength, there are three distinct regions where the linear trend can be observed. Further, it also shows that the region boundaries are not fixed but change with system parameters. In the following analysis, the piecewise linear approximation is derived by solving~\eqref{eq:28} under different operation regimes based on the self-interference signal strength.
\subsubsection{Strong self-interference regime}
In this case the received self-interference signal strength is assumed to be strong enough such that the noise introduced due to the presence of the self-interference signal ($\eta \mathsf{RSSI}_A$) is higher than the receiver noise floor i.e. $\eta \mathsf{RSSI}_A > \zeta$ or $\mathsf{RSSI}_A > \frac{\zeta}{\eta}$. Moreover, $\eta$ is a combination of phase, receiver, and quantization noises; in practical fabricated circuits, all of these noise components are typically $\ll 1$ (see~\cite{Ref15}, and~\cite{Ref15r})\footnote{based on the data sheets' numbers in~\cite{Ref15, Ref15r}, the total inband phase noise $\mu$ in a 1MHz bandwidth is $\sim-$40dB, the total ADC quantization noise $\sigma_q^2$=-77dB, and the mixer noise figure $N_m$=10dB. By substituting in~\eqref{eq:13} we get $\eta\simeq$39.9dB (i.e. $\sim 1e^{-4}$) which is $\ll$ 1}. Accordingly, assuming that $\mathsf{RSSI}_A > \frac{\zeta}{\eta}$, and $\eta \ll 1$, Equation~\eqref{eq:24} can be approximated to
\begin{equation}\label{eq:29}
a \cong \eta\text{, } b \cong \frac{\zeta}{2}\text{, } c \cong -(\eta \mathsf{RSSI}_A)^2\text{.}
\end{equation}
Substituting from~\eqref{eq:29} in~\eqref{eq:28}, we get
\begin{equation}\label{eq:32}
\mathsf{RSSI}_{B,min} \cong \frac{\frac{-\zeta}{2}+\sqrt{\frac{\zeta^2}{4}+4\eta^3 \mathsf{RSSI}_A^2}}{2\eta}\text{.}
\end{equation}

First, we study the case where $4\eta^3 \mathsf{RSSI}_A^2 > \frac{\zeta^2}{4}$ i.e. $\mathsf{RSSI}_A > \frac{\zeta}{4\eta \sqrt{\eta}}$. Using Taylor expansion of $(1+x)^{\frac{1}{2}} \cong 1+\frac{1}{2} x$ when $x<1$, Equation~\eqref{eq:32} can be approximated as
\begin{eqnarray}\label{eq:33}
\mathsf{RSSI}_{B,min} &\cong& \frac{-\frac{\zeta}{2} + 2\eta \sqrt{\eta} \mathsf{RSSI}_A \sqrt{1 + \frac{\zeta^2}{16\eta^3 \mathsf{RSSI}_A^2}}}{2\eta} \nonumber \\       
&\cong& \frac{-\frac{\zeta}{2} + 2\eta \sqrt{\eta} \mathsf{RSSI}_A + \frac{\zeta^2}{16\eta \sqrt{\eta} \mathsf{RSSI}_A}}{2\eta}\text{.}
\end{eqnarray}
Using the case condition $\mathsf{RSSI}_A > \frac{\zeta}{4\eta \sqrt{\eta}}$, equation~\eqref{eq:33} can be approximated as
\begin{equation}\label{eq:34}
\mathsf{RSSI}_{B,min} \cong \frac{\eta \mathsf{RSSI}_A}{\sqrt{\eta}}\text{.}
\end{equation}

Now, considering the opposite case where $4\eta^3 \mathsf{RSSI}_A^2 < \frac{\zeta^2}{4}$ i.e. $\mathsf{RSSI}_A < \frac{\zeta}{4\eta \sqrt{\eta}}$, Equation~\eqref{eq:32} can be approximated as
\begin{eqnarray}\label{eq:35}
\mathsf{RSSI}_{B,min} &\cong& \frac{-\frac{\zeta}{2} + \frac{\zeta}{2}\sqrt{1+\frac{16\eta^3 \mathsf{RSSI}_A^2}{\zeta^2}}}{2\eta} \nonumber \\
&\cong& \frac{2\eta^2 \mathsf{RSSI}_A^2}{\zeta}\text{,}
\end{eqnarray}
where $\frac{\zeta}{\eta} < \mathsf{RSSI}_A < \frac{\zeta}{4\eta \sqrt{\eta}}$.
\subsubsection{Weak self-interference regime}

In this case the received self-interference signal strength is assumed to be weak enough such that the noise introduced due to the presence of the self-interference signal ($\eta \mathsf{RSSI}_A$) is lower than the receiver noise floor i.e. $\eta \mathsf{RSSI}_A < \zeta$ or $\mathsf{RSSI}_A < \frac{\zeta}{\eta}$. Assuming that $\mathsf{RSSI}_A < \frac{\zeta}{\eta}$, and $\eta \ll 1$, Equation~\eqref{eq:24} can be approximated to
\begin{equation}\label{eq:36}
a \cong \eta\text{, } b \cong \frac{\zeta}{2}\text{, } c \cong -\zeta \eta \mathsf{RSSI}_A\text{.}
\end{equation}
Substituting from~\eqref{eq:36} in~\eqref{eq:28}, we get
\begin{equation}\label{eq:37}
\mathsf{RSSI}_{B,min} \cong \frac{-\frac{\zeta}{2} + \sqrt{\frac{\zeta^2}{4}+4\zeta \eta^2 \mathsf{RSSI}_A}}{2\eta}\text{.}
\end{equation}
For additional simplification, first, assume that $4\zeta \eta^2 \mathsf{RSSI}_A > \frac{\zeta^2}{4}$ i.e. $\mathsf{RSSI}_A > \frac{\zeta}{16\eta^2}$. Knowing that typically $\eta \ll 1$, this condition contradicts the weak self-interference condition above, and thus $4\zeta \eta^2 \mathsf{RSSI}_A$ should be $< \frac{\zeta^2}{4}$. Accordingly, equation~\eqref{eq:37} can be approximated as
\begin{eqnarray}\label{eq:38}
\mathsf{RSSI}_{B,min} &\cong& \frac{-\frac{\zeta}{2} + \frac{\zeta}{2}\sqrt{1+\frac{16\eta^2 \mathsf{RSSI}_A}{\zeta}}}{2\eta} \nonumber \\
&\cong& 2\eta \mathsf{RSSI}_A\text{,}
\end{eqnarray}
where $\mathsf{RSSI}_A < \text{min}(\frac{\zeta}{\eta},\frac{\zeta}{16\eta^2})$. Since $\eta \ll 1$ for practical systems, it can be assumed that $16\eta^2 < \eta$. Therefore, the condition for this operation regime is $\mathsf{RSSI}_A < \frac{\zeta}{\eta}$

As a conclusion, using~\eqref{eq:34},~\eqref{eq:35}, and~\eqref{eq:38}, the simplified rate gain region for digital cancellation scheme can be written as
\begin{equation}\label{eq:39}
\mathsf{RSSI}_{B,min} \cong \left\{
\begin{array}{c l}
    \frac{\eta \mathsf{RSSI}_A}{\sqrt{\eta}} & \text{,\ \ } \mathsf{RSSI}_A \geq \frac{\zeta}{4\eta \sqrt{\eta}}\text{,}\\
    \frac{2\eta^2 \mathsf{RSSI}_A^2}{\zeta} & \text{,\ \ } \frac{\zeta}{\eta} \leq \mathsf{RSSI}_A < \frac{\zeta}{4\eta \sqrt{\eta}}\text{,}\\
	 2\eta \mathsf{RSSI}_A & \text{,\ \ } \mathsf{RSSI}_A < \frac{\zeta}{\eta}\text{.}
\end{array}\right.
\end{equation}
\subsection{Rate gain region for analog cancellation scheme \label{sec:analograte}}
Due to the similarity of the SINR relations in both digital and analog cancellation schemes, the rate gain region for analog cancellation acheme can be derived following the same steps as in Section~\ref{sec:digitalrate} above by using Equation~\eqref{eq:12} instead of~\eqref{eq:11}. Thus, for simplicity, we present the final results without going into derivation details. Equation~\eqref{eq:40} describes the piecewise linear approximation for the rate gain region in analog cancellation scheme.
\begin{equation}\label{eq:40}
\mathsf{RSSI}_{B,min} \cong \left\{
\begin{array}{c l}
    \frac{\mu \mathsf{RSSI}_A}{\sqrt{\eta}} & \text{,\ \ } \mathsf{RSSI}_A \geq \frac{\zeta}{4\mu \sqrt{\mu}}\text{,}\\
    \frac{2\mu^2 \mathsf{RSSI}_A^2}{\zeta} & \text{,\ \ } \frac{\zeta}{\eta} \leq \mathsf{RSSI}_A < \frac{\zeta}{4\mu \sqrt{\mu}}\text{,}\\
	 2\mu \mathsf{RSSI}_A & \text{,\ \ } \mathsf{RSSI}_A < \frac{\zeta}{\mu}\text{.}
\end{array}\right.
\end{equation}

\begin{figure}[!ht]
\begin{center}
\noindent
  \includegraphics[width=6.5in,trim= 0in 0in 0in 0in]{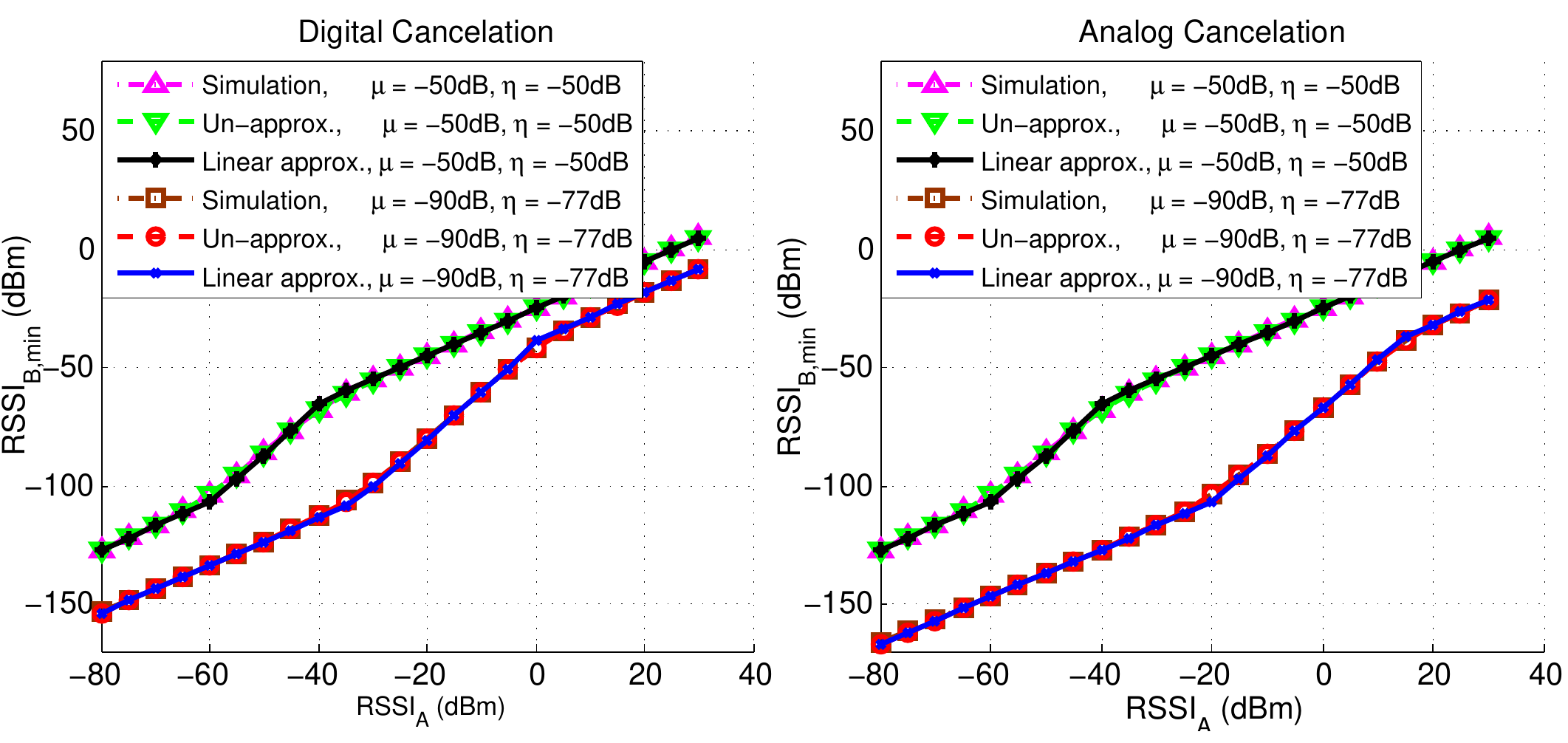}
  \caption{Comparison between simulation and analytical results for digital and analog cancellation scheme.\label{Fig4Label}}
\end{center}
\end{figure}

Using~\eqref{eq:39} and~\eqref{eq:40}, one can straightforwardly predict the region where full-duplex systems outperform half-duplex systems for any given combination of system parameters and operating conditions. The accuracy of the approximated rate gain region described by~\eqref{eq:39} and~\eqref{eq:40} is confirmed by comparing it to the un-approximated rate gain region at different system parameters. Furthermore, the accuracy of the analysis is verified by comparing both the approximated and the un-approximated rate gain regions to the simulation results. The comparison results are shown in Figures~\ref{Fig4Label}. It is clear that the results from simulation closely match the un-approximated analysis, which validates the accuracy of the analysis. In addition, the results also show that the approximated rate gain region is an excellent fit to the un-approximated one except at the transition between different operation regions where a small error (0 - 3dBm) exists.

\section{Key observations and design tradeoffs}
In this section, we highlight several key observations regarding full-duplex system behaviour under analog and digital self-interference cancellation schemes.

Observation 1: \emph{Analog cancellation reduces the effect of both mixer and quantization noise}. According to~\eqref{eq:39} and~\eqref{eq:40}, the rate gain region for the digital cancellation scheme depends on the combined noise power associated with the self-interference signal ($\eta \mathsf{RSSI}_A$) which consists of all noise components. However, in the analog cancellation scheme, the rate gain region only depends on the phase noise power ($\mu \mathsf{RSSI}_A$). This observation implies that using the analog cancellation scheme reduces the effect of all noise components except phase noise. The reason is that analog cancellation eliminates most of the self-interference power at the LNA input allowing it to operate at a high gain mode, and thus reducing the effect of both mixer and quantization noise (review ~\eqref{eq:7} and ~\eqref{eq:10}). On the other hand, changing the LNA gain does not affect the phase noise associated with the incoming self-interference signal. 

Observation 2: \emph{Analog cancellation is most useful for low phase noise systems}. As a consequence of the previous observation, when phase noise dominates other noise components (i.e. $\mu \gg \sigma_q^2+P_{th} N_m$, and therefore $\eta \cong \mu $), the rate gain region will be identical for both digital and analog cancellation schemes. On the other hand, if $\mu \ll \sigma_q^2+P_{th} N_m$ (i.e. either quantization or mixer noise dominates), the analog cancellation scheme outperforms the digital cancellation scheme. Therefore, the advantage of using analog cancellation is evident only when either quantization or mixer noise dominates the phase noise. Accordingly, in high phase noise systems, performing analog cancellation requires additional hardware complexity~\cite{Ref5} without achieving performance gain over digital cancellation scheme.

Looking at Observations 1 and 2 one can conclude that current full-duplex systems are limited by the oscillator phase noise. From a noise perspective, the results in~\eqref{eq:39} and~\eqref{eq:40} shows that, for analog cancellation scheme, the rate gain region is limited by the total phase noise $\mu$. While, for digital cancellation scheme, the rate gain region is limited by the combined noise parameter $\eta$, which consists of all phase, quantization, and mixer noise. Therefore, the system bottleneck for analog cancellation scheme is the phase noise. 

On the other hand, in the digital cancellation scheme, the bottleneck is the dominant component of phase, quantization, and mixer noise. Typically, in today's wireless technology, down-conversion mixer's noise figure is $\sim$10dB~\cite{Ref15r} resulting in a normalized mixer noise power ($P_{th} N_m$) of $-$104dBm in a 1MHz bandwidth. Further, assuming a 12 bits ADC is used, the resulting normalized quantization noise power is $\sim$ $-$77dBm. However, the in-band oscillator phase noise is usually much higher than those values, for example, the 2.4GHz oscillators in~\cite{Ref17} has an total in-band phase noise of $-$50dBc in a 1MHz bandwidth. Thus the in-band oscillator phase noise dominates other noise components, and is considered the bottleneck for current full-duplex systems with either analog or digital cancellation schemes.

Observation 3: \emph{Significant performance improvement requires both passive suppression and hardware enhancement}. According to~\eqref{eq:39} and~\eqref{eq:40}, improving the rate gain region could be achieved through one/both of two main techniques, i) reducing the received self-interference signal strength and/or  ii) improving analog circuits to reduce noise. Each technique has tradeoffs that might limit its applicability and practicality. For example, one way to reduce self-interference RSSI is to passively suppress the self-interference in the spatial domain before it is processed by the receiver radio-frequency section. However, passive self-interference suppression schemes (e.g. antenna separation~\cite{Ref5}, antenna cancellation~\cite{Ref6}, and antenna directionality~\cite{Ref9}) usually have a limited mitigation capability. The second improvement technique is to reduce the noise introduced by the analog front-end through either technology, device or architectural innovations. From a practical viewpoint, noise reduction in analog circuits is very challenging, and the improvement could be very limited. Therefore, achieving wide rate gain region require a hybrid approach that combines contributions from both techniques. 

The first tradeoff: \emph{Increasing the rate gain region at the cost of hardware complexity}. According to observation 1 and 2, in some cases when either mixer or quantization noise dominates the phase noise, performing analog cancellation reduces the noise effect and improves the overall system performance. However, this performance enhancement comes at the cost of additional hardware required to perform analog cancellation~\cite{Ref5,Ref7}.

The second tradeoff: \emph{Increasing the rate gain region by reducing the transmission range}. According to~\eqref{eq:39} and~\eqref{eq:40}, reducing the received self-interference signal strength increases the rate gain region. One way to reduce self-interference RSSI is to reduce the transmitted signal power. However, reducing the transmit power also reduces the signal-of-interest power at the desired node, which in effect reduces the transmission range. From a practical point of view, improving the rate gain region by trading the transmission range might be beneficial for short range applications or for applications where symmetrical transmit and receive rates are not necessary. 

The third tradeoff: \emph{Increasing the passive suppression to allow for higher noise levels}. Equation~\eqref{eq:39} and~\eqref{eq:40} show that for same design target (rate gain region), there exists a tradeoff between noise reduction and passive self-interference suppression. In other words, additional passive self-interference suppression allows for having higher noise values while maintaining the same performance and vice versa. For example, for analog cancellation scheme, Equation~\eqref{eq:40} shows that there is a linear relation (in the log-domain) with a slope of one between the self-interference RSSI and the phase noise required to achieve certain performance. Accordingly, an $x$dB additional passive self-interference suppression allows the phase noise to be higher by the same $x$ amount while achieving the same performance.

\section{Numerical Results}
In this section we numerically investigate the full-duplex system performance, design tradeoffs, and rate gain region for practical indoor applications. First, we use the rate gain region to investigate the design requirements to enable full-duplex transmission with rate gains as compared to half-duplex transmission. Then, we characterize the rate gain that could be achieved by using full-duplex instead of half-duplex as a communication scheme. In this analysis, the cluster-based channel model introduced in~\cite{Ref19} is used to model the wireless channel.
%
%
The signal-of-interest and self-interference channel Rician factors are chosen according to the experimental results presented in~\cite{Ref20} to be 0dB and 35dB respectively. The propagation loss (in units of dB) is assumed to follow the log-normal model with shadowing effect, and can be written as~\cite{Ref21}
\begin{equation}\label{eq:43}
L = -K + 10\text{\ }r\text{\ } \log(d)+X_{\sigma}\text{,}
\end{equation}
where $K=10log(\frac{\lambda}{4\pi})^2$ and $\lambda$ is the carrier wavelength, $r$ is the propagation loss exponent, $d$ is the distance traveled by the signal, and $X \sim \mathcal{N}(0,\sigma^2)$ represents the shadowing effect. According to the empirical results in~\cite{Ref21} $r\cong$ 2.5, and $\sigma \cong$ 3.5dB for practical indoor environments with LOS component. System parameters are chosen to reflect industry standard chipsets~\cite{Ref15,Ref15r} operating in the ISM band as follows: the carrier frequency $f_c=$2.4GHz, system $BW=$1MHz, LNA noise figure $N_l=$4dB, mixer noise figure $N_m=$10dB, and number of ADC bits $m=$12bits.
\subsection{Design requirements for feasible full-duplex transmission}
Achieving higher full-duplex rate gain over larger range requires full-duplex systems to have a wide rate gain region such that the received signal strength falls within the rate gain region. According to~\eqref{eq:39} and~\eqref{eq:40}, the full-duplex rate gain region is mainly controlled by the system noise level and the received self-interference signal strength. The received self-interference signal strength could be written as a multiplication of the transmit power and the passive self-interference suppression as $\mathsf{RSSI}_A = P_x C$, where $C$ is the passive self-interference suppression due to antenna separation and/or other passive suppression techniques. The rate gain region becomes a function of three main parameters: the noise level, the transmit power, and the passive suppression. In fact, to achieve a certain rate gain region, different combinations could be used.

\begin{figure}[!ht]
\begin{center}
\noindent
  \includegraphics[width=6in,trim= 0in 0in 0in 0in]{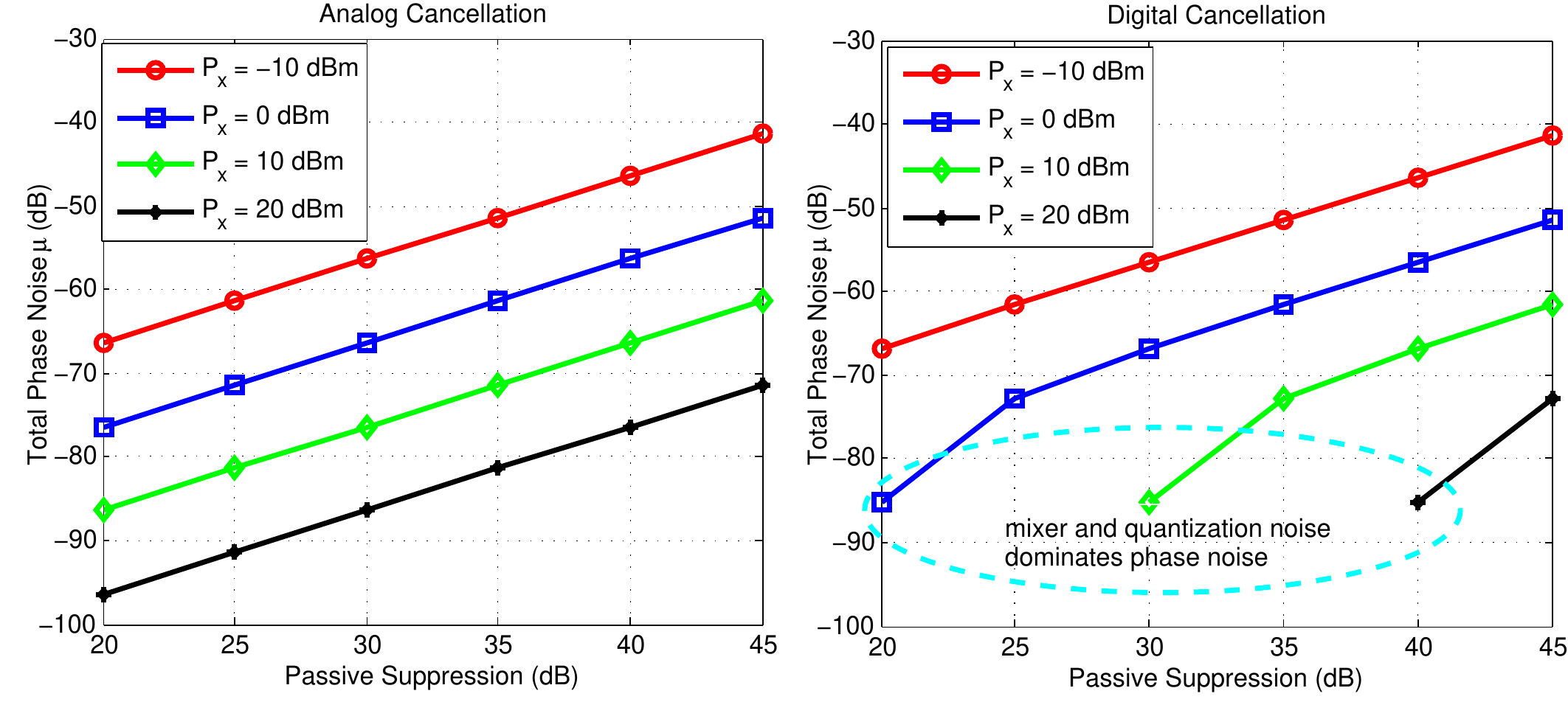}
  \caption{Requirements on phase noise level and passive self-interference suppression for rate gain region of $-$80dBm in case of analog and digital cancellation schemes.\label{Fig67Label}}
\end{center}
\end{figure}

As a design example, we quantify the requirements for full-duplex systems to achieve better rate than half-duplex systems in practical indoor applications such as Bluetooth. According to the experimental results in~\cite{Ref23}, the average received signal strength for typical Bluetooth signal is in the range of $-$65dBm to $-$80dBm. Therefore, choosing a design target of $\mathsf{RSSI}_{B,min}=$ $-$80dBm guarantees that most of the incoming signal power in such application falls within the rate gain region. Figure~\ref{Fig67Label} shows the combination of the total phase noise ($\mu$), transmit power, and passive suppression required to achieve $-$80dBm rate gain region for both analog and digital cancellation schemes.
%
%
%
%

The results illustrate the tradeoffs discussed in Section IV. It shows that at high phase noise values (when phase noise dominates other noises), analog and digital cancellation schemes achieve the same performance. While at low phase noise values, the analog cancellation scheme outperforms the digital cancellation scheme. For example, at phase noise of $-$85dB and transmit power of 10dBm, the analog cancellation scheme requires 8dB less passive suppression than digital cancellation. The results also show the tradeoff between the transmit power, the system noise, and the passive suppression. The exact relation between the three parameters could be derived using~\eqref{eq:39},~\eqref{eq:40} along with the assumed system parameters at the beginning of this section. It can be shown that for a rate gain region of $-$80dBm, the relation between the transmit power, the noise level and the amount of passive suppression for both digital and analog cancellation schemes respectively can be written as
\begin{equation}\label{eq:44}
P_x + \eta + C = -96.5,
\end{equation}
\begin{equation}\label{eq:45}
P_x + \mu + C = -96.5,
\end{equation}
where all parameters are in dB units. Equation~\eqref{eq:44} and~\eqref{eq:45} show that the transmit power, the noise level and the passive suppression could be traded with each other to achieve the same design target. For example, we can trade some additional passive suppression (that could be achieved by increasing the antenna separation implying larger device sizes) with more transmit power (i.e. more transmission range) or having higher noise (i.e. less hardware complexity). We could also lower the transmit power (i.e. less transmission range) while trading the passive suppression (i.e. smaller devices) and the noise level.

To better articulate these observations, we consider the example of Bluetooth system as a practical application and study different design tradeoffs. The three Bluetooth system's classes are considered in this analysis~\cite{Ref24}. The Bluetooth application is operating in the 2.4GHz band, thus the oscillator in~\cite{Ref17}, which has a $\sim-$50dBc total in-band phase noise ($\mu$) is assumed.

In class-3 Bluetooth systems, the transmit power is 0dBm and achieves $\sim$1 meters transmission range. Substituting in~\eqref{eq:45}, we find that a $\sim$46dB passive self-interference suppression is required to achieve $-$80dBm rate gain region. For wider transmission range (e.g. $\sim$10 meters), class-2 Bluetooth could be used. However, the transmit power in this case is 4dBm, which means 4dB more passive suppression is required. In class-1 Bluetooth systems, a much wider transmission range of $\sim$100 meters could be achieved at transmit power of 20dBm. However, in this case, a $\sim$66dB passive suppression is required. This example illustrates the tradeoff between the transmission range and the amount of suppression cancellation using an example of current radio circuits. However, a possible means to avoid the aggressive requirements on the amount of passive suppression is to trade it with the oscillator phase noise. According to~\eqref{eq:45}, each $x$dB reduction in the oscillator in-band phase noise reduces the requirements on the passive suppression by the same amount, thus promoting aggressive in-band noise reduction techniques for oscillator circuits.

\subsection{Achievable Rate gain}
\begin{figure}[!ht]
\begin{center}
\noindent
  \includegraphics[width=6in,trim= 0in 0in 0in 0in]{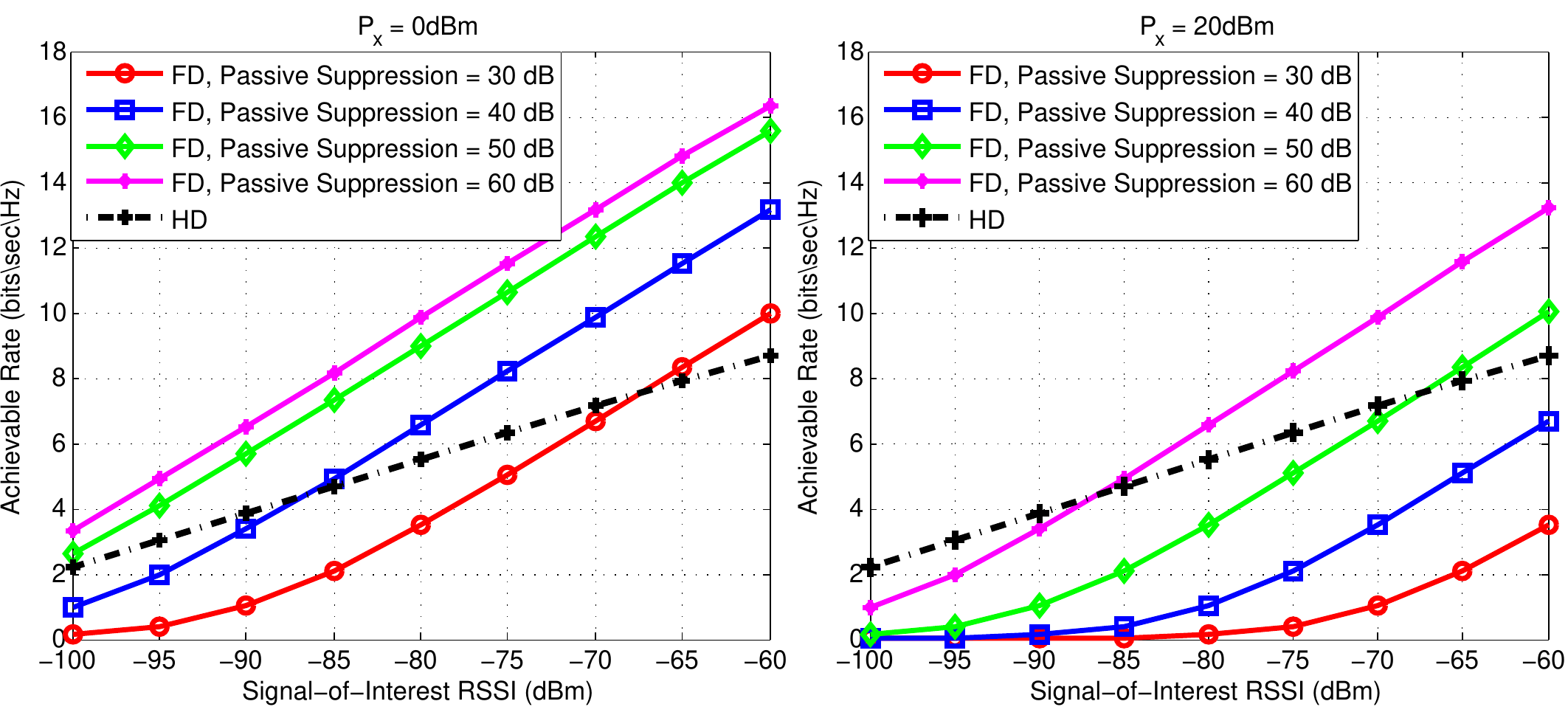}
  \caption{Achievable rate for full-duplex and half-duplex systems with total phase noise $\mu$=$-$60dB, and transmit power of 0dBm and 20dBm.\label{Fig89Label}}
\end{center}
\end{figure}

Another important aspect to investigate is the improvement in spectral efficiency when full-duplex transmission is employed by evaluating the achievable rate gain at different signal-of-interest RSSI values. Figure~\ref{Fig89Label} shows the achievable rate for both full-duplex and half-duplex systems at different passive self-interference suppression amounts with transmit power of 0dBm and 20dBm. The results show that, at 0dBm transmit power, a total of 40dB passive self-interference suppression could achieve $-$87dBm rate gain region, while achieving a rate of $\sim$1.2x and $\sim$1.4x times the half-duplex rate at $-$80dBm and $-$70dBm signal-of-interest strengths respectively, which is considered a significant improvement in the system throughput. The results also show that, at 20dBm transmit power, the full-duplex system requires 20dB more passive suppression to achieve the same performance as the case of 0dBm transmit power, which is consistent with the results in~\eqref{eq:45}.

On the other hand, the half-duplex rate is identical in both simulation cases. The reason is that, half-duplex system's performance depends on the received signal-of-interest strength, which is the same in both simulation cases. In this specific simulation, in order to keep the same received signal-of-interest strength, the transmission range (distance between the two communicating nodes) is increased along with the increase of the transmit power.

\begin{figure}[!ht]
\begin{center}
\noindent
  \includegraphics[width=3.5in,trim= 0in 0in 0in 0in]{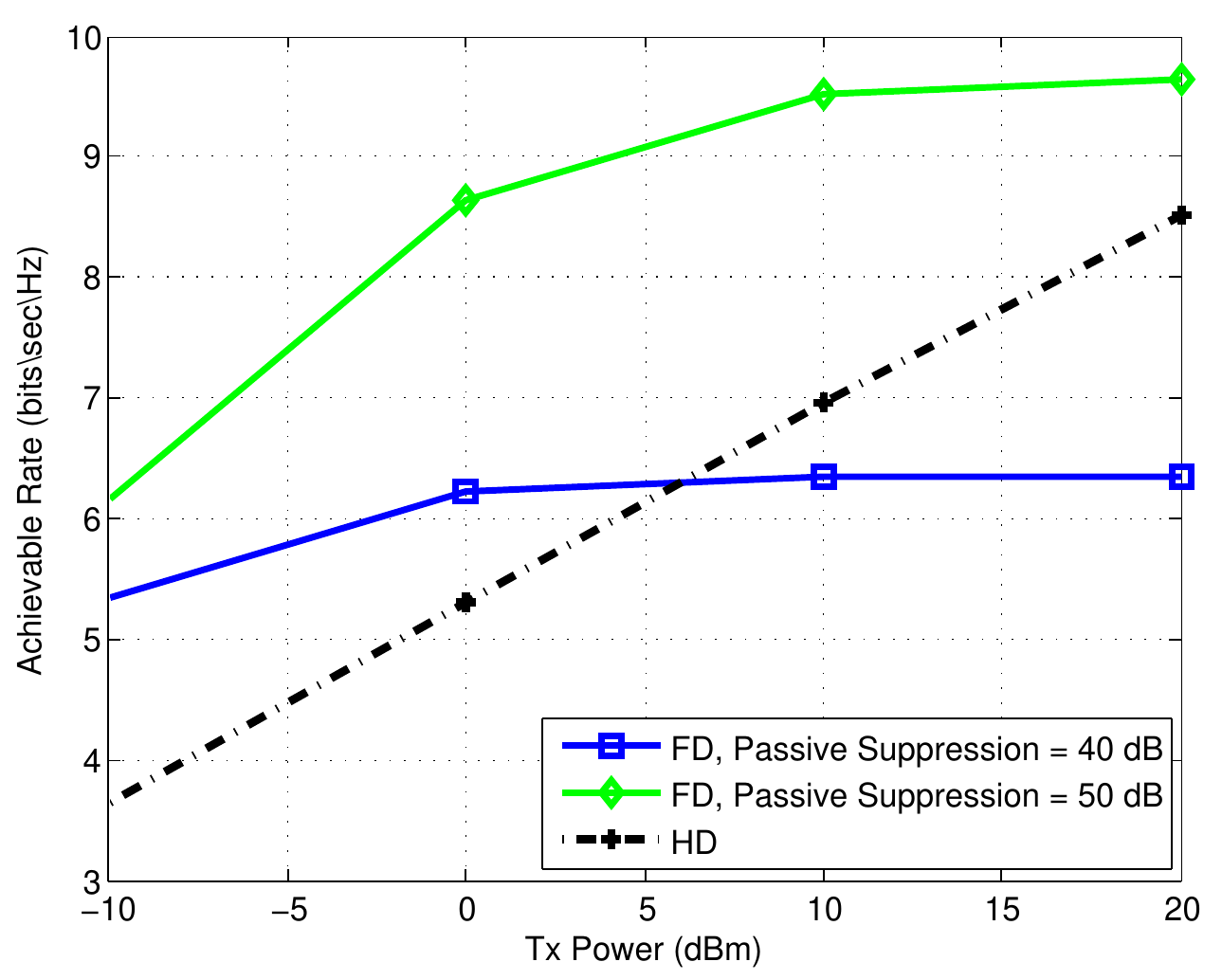}
  \caption{Achievable rate for full-duplex and half-duplex systems at different transmit power values, with total phase noise $\mu$=$-$60dB, and distance $D=50$ meters between the two communicating nodes.\label{Fig10Label}}
\end{center}
\end{figure}

Figure~\ref{Fig89Label} implicitly show that, increasing the transmit power worsens the full-duplex system performance. In the following simulation, we numerically investigate the effect of changing the transmit power (while keeping constant distance($D$) between the two communicating nodes) on both full-duplex and half-duplex system's performance. Figure~\ref{Fig10Label} shows the full-duplex and half-duplex achievable rate at different transmit power values. The results show that, the full-duplex rate gain decreases with the increase of the transmit power, which implies that, for a given distance $D$, lowering the transmit power makes full-duplex systems more likely to outperform half-duplex systems. This conclusion consists with the second tradeoff discussed in section IV. The results also show that, as the transmit power increases the full-duplex rate increases until it reaches a saturation point. The reason is that, as the transmit power increases, the received self-interference signal strength increases, and the full-duplex SINR starts to be totally limited by the un-cancelled self-interference power ($\eta\mathsf{RSSI}_A$). At this point, increasing the transmit power will increase both signal-of-interest and un-cancelled self-interference power with the same amount, keeping the SINR constant.

\section{Conclusion}
In this paper we introduced a signal model for single input single output narrowband full-duplex system by modeling different transmitter and receiver radio impairments. More specifically, transmitter and receiver phase noise, LNA noise figure, mixer noise figure, and analog-to-digital converter quantization noise. The signal model is used to analytically derive a piecewise linear, in the log domain, approximation for the rate gain region in terms of all system parameters as well as radio impairments under both analog and digital self-interference cancellation schemes. A study of full duplex system behaviour under different operation conditions is presented illustrating the system design space and possible tradeoffs.  Finally, we numerically investigate the design requirements to enable full-duplex transmission with rate gains as compared to half-duplex transmission in typical indoor environments. The results show that, for short range applications such as class-3 Bluetooth, a $-$60dBc in-band phase noise combined with 40dB passive self-interference suppression could achieve a rate of $\sim$1.2x to $\sim$1.4x times that of half-duplex systems. On the other hand, for longer range applications such as class-1 Bluetooth, $-$70dB in-band phase noise combined with 50dB passive self-interference suppression is required to achieve the same rate gain.



\begin{thebibliography}{1}

%

\bibitem{Ref3}
A. Sahai, G. Patel, and A. Sabharwal, ``Asynchronous full-duplex wireless," \emph{Communication Systems and Networks (COMSNETS), 2012 Fourth International Conference on} , pp.1-9, Jan. 2012.

\bibitem{Ref4}
S. Li, and R. D. Murch, ``Full-Duplex Wireless Communication Using Transmitter Output Based Echo Cancellation," \emph{Global Telecommunications Conference (GLOBECOM 2011)}, 2011 IEEE , pp.1-5, Dec. 2011.

\bibitem{Ref5}
M. Duarte, and A. Sabharwal, ``Full-duplex wireless communications using off-the-shelf radios: Feasibility and first results," \emph{Signals, Systems and Computers (ASILOMAR), 2010 Conference Record of the Forty Fourth Asilomar Conference on}, pp.1558-1562, Nov. 2010.

\bibitem{Ref6}
J. I. Choi, M. Jain, K. Srinivasan, P. Levis, and S. Katti, ``Achieving single channel, full duplex wireless communication," \emph{in MobiCom}, 2010.

\bibitem{Ref7}
B. Radunovic, D. Gunawardena, P. Key, A. P. N. Singh, V. Balan, and G. Dejean, ``Rethinking indoor wireless Mesh Design: Low power, low frequency, full duplex," \emph{Wireless Mesh Networks (WIMESH 2010), 2010 Fifth IEEE Workshop on}, pp.1-6, June 2010.

\bibitem{Ref8}
E. Everett, M. Duarte, C. Dick, and A. Sabharwal, ``Empowering full-duplex wireless communication by exploiting directional diversity," \emph{Signals, Systems and Computers (ASILOMAR), 2011 Conference Record of the Forty Fifth Asilomar Conference on}, pp.2002-2006, Nov. 2011.

\bibitem{Ref9}
E. Ahmed, A. M. Eltawil, and A. Sabharwal, ``Simultaneous Transmit and Sense for Cognitive Radios Using Full-duplex: A First Study," to appear: \emph{2012 IEEE International Symposium on Antennas and Propagation and USNC-URSI National Radio Science Meeting}.

\bibitem{Ref9r}
M. Duarte, A. Sabharwal, V. Aggarwal, R. Jana, K. K. Ramakrishnan, C. Rice, and N. K. Shankaranarayanan, ``Design and Characterization of a Full-duplex Multi-antenna System for WiFi networks," submitted to \emph{IEEE Transactions on Vehicular Technology}, Oct. 2012. [Online]. Available: http://arxiv.org/abs/1210.1639.

\bibitem{Ref9rr}
W. Cheng, X. Zhang, and H. Zhang, ``Full Duplex Wireless Communications for Cognitive Radio Networks," [Online]. Available: http://arxiv.org/abs/1105.0034.

\bibitem{Ref10}
P. Lioliou, M. Viberg, M. Coldrey, and F. Athley, ``Self-interference suppression in full-duplex MIMO relays," \emph{Signals, Systems and Computers (ASILOMAR), 2010 Conference Record of the Forty Fourth Asilomar Conference on} , pp.658-662, Nov. 2010.

\bibitem{Ref10r}
T. Kwon, S. Lim, S. Choi, and D. Hong, ``Optimal Duplex Mode for DF Relay in Terms of the Outage Probability," \emph{Vehicular Technology, IEEE Transactions on}, vol.59, no.7, pp.3628-3634, Sept. 2010.

\bibitem{Ref10rr}
B. P. Day, A. R. Margetts, D. W. Bliss, and P. Schniter, ``Full-Duplex MIMO Relaying: Achievable Rates Under Limited Dynamic Range," \emph{Selected Areas in Communications, IEEE Journal on}, vol.30, no.8, pp.1541-1553, September 2012.

\bibitem{Ref11}
B. P. Day, A. R. Margetts, D. W. Bliss, and P. Schniter, ``Full-Duplex Bidirectional MIMO: Achievable Rates Under Limited Dynamic Range," \emph{Signal Processing, IEEE Transactions on}, vol.60, no.7, pp.3702-3713, July 2012.

\bibitem{Ref13}
B. Razavi, Design of Analog CMOS Integrated Circuits. New York:McGraw-Hill, 2001.

\bibitem{Ref14}
M. Lu, N. Shanbhag, and A. Singer, ``BER-optimal analog-to-digital converters for communication links," \emph{Circuits and Systems (ISCAS), Proceedings of 2010 IEEE International Symposium on}, pp.1029-1032, May 30 2010-June 2 2010.

\bibitem{Ref12}
A. Mehrotra, ``Noise analysis of phase-locked loops," \emph{Circuits and Systems I: Fundamental Theory and Applications, IEEE Transactions on}, vol.49, no.9, pp. 1309-1316, Sep 2002.

\bibitem{Ref15}
\url{``http://datasheets.maximintegrated.com/en/ds/MAX2828-MAX2829.pdf''}.

\bibitem{Ref15r}
\url{``http://datasheets.maximintegrated.com/en/ds/MAX19757.pdf''}.


\bibitem{Ref17}
S. Shekhar, D. Gangopadhyay, W. Eum-Chan, and D. J. Allstot, ``A 2.4-GHz Extended-Range Type-I $\Sigma\Delta$  Fractional-$N$ Synthesizer With 1.8-MHz Loop Bandwidth and 110-dBc/Hz Phase Noise," \emph{Circuits and Systems II: Express Briefs, IEEE Transactions on}, vol.58, no.8, pp.472-476, Aug. 2011.

\bibitem{Ref19}
V. Erceg, and L. Schumacher et al., ``TGn channel models," IEEE 802.11-03/940r4, May 2004.

\bibitem{Ref20}
M. Duarte, C. Dick, and A. Sabharwal, ``Experiment-driven Characterization of Full-Duplex Wireless Systems," to appear in \emph{IEEE Transactions on Wireless Communications}. [Online]. Available: arXiv:1107.1276v2.

\bibitem{Ref21}
S. S. Ghassemzadeh, L. J. Greenstein, A. Kavcic, T. Sveinsson, and V. Tarokh, ``UWB indoor path loss model for residential and commercial buildings," \emph{Vehicular Technology Conference, 2003. VTC 2003-Fall. 2003 IEEE 58th}, vol.5, pp.3115-3119 Vol.5, Oct. 2003.


\bibitem{Ref23}
S. Zhou, and J. K. Pollard, ``Position measurement using Bluetooth," \emph{Consumer Electronics, IEEE Transactions on}, vol.52, no.2, pp.555-558, May 2006.

\bibitem{Ref24}
SIG, ``Specification of the Bluetooth System", http://www.bluetooth.com.


\end{thebibliography}
\end{document}